# Image encryption for Offshore wind power based on 2D-LCLM and Zhou Yi Eight Trigrams

## Lei Kou, Jinbo Wu, Fangfang Zhang* and Peng Ji

Qilu University of Technology (Shandong Academy of Sciences), Qingdao, China

*Corresponding author

## Wende Ke

Southern University of Science and Technology, Shenzhen, China

## Junhe Wan and Hailin Liu

Qilu University of Technology (Shandong Academy of Sciences), Qingdao, China

## Yang Li

Northeast Electric Power University, Jilin, China

## Quande Yuan

Changchun Institute of Technology, Changchun, China

**Abstract:** Offshore wind power is an important part of the new power system, due to the complex and changing situation at ocean, its normal operation and maintenance cannot be done without information such as images, therefore, it is especially important to transmit the correct image in the process of information transmission. In this paper, we propose a new encryption algorithm for offshore wind power based on two-dimensional lagged complex logistic mapping (2D-LCLM) and Zhou Yi Eight Trigrams. Firstly, the initial value of the 2D-LCLM is constructed by the Sha-256 to associate the 2D-LCLM with the plaintext. Secondly, a new encryption rule is proposed from the Zhou Yi Eight Trigrams to obfuscate the pixel values and generate the round key. Then, 2D-LCLM is combined with the Zigzag to form an S-box. Finally, the simulation experiment of the algorithm is accomplished. The experimental results demonstrate that the algorithm can resistant common attacks and has prefect encryption performance.

## 1 Introduction

With the growing demand for electricity, the traditional power system is facing increasing problems such as resource shortage and environmental pollution [1]. Offshore wind power is becoming a new direction for renewable energy due to its cleanliness, pollution-free and abundant resources [2]. However, limited by the natural environment of the ocean, offshore wind power relies on pictures and other information for operation and maintenance, which leads to another problem: how to secure image during transmission.

Using traditional text encryption methods has low capability in confusing and diffusing image [3], therefore, many scholars combine other disciplines to design excellent image encryption algorithms [4-6]. Cryptography and chaotic systems coincide in many properties such as initial value sensitivity, randomness, convenience, and uncertainty [7]. Based on these properties of chaotic systems, they can be applied to image encryption for pixel confusion and diffusion, etc. Therefore, numerous researchers have combined chaos and image encryption to propose a large number of encryption algorithms [8,9].

Compared with chaotic systems, complex chaotic systems have more excellent properties, such as more complex behaviors and higher dimensionality which have led more and more scholars to favor the study of complex chaos [10]. Zhang et al. proposed a new 9D complex chaotic system and used it in encryption of smart grid images [11]. Zhang et al. analyzed the characteristics of two-dimensional lagged complex logistic mapping (2D-LCLM) and applied it to image encryption [12].

The terminal devices for collecting data in offshore wind farms are limited by the resource constraints, their storage space is not very large and their processing power for data is relatively weak. Therefore, it is good to use low-power, highly confusing lightweight ciphers when encrypting and decrypting images [13]. The implementation of lightweight cipher mainly includes S-boxes, and the same S-box in serial can well reduce the resources required to implement the algorithm [14,15]. Çavuşoğlu et al. proposed an S-box with strong cryptographic properties based on a genetic algorithm [16]. Hua et al. constructed a new S-box and used it in image encryption, which experimentally proved to be a great improvement in the security of the algorithm [17].

The Zhou Yi Eight Trigrams is an excellent ancient Chinese culture. Zhang et al. designed a new segmentation method based on the I-Ching operator, which effectively suppresses image noise [18]. The Zhou Yi Eight Trigrams contains many variations, which can be used to disrupt the relationship between pixels to achieve encryption. In addition, Zhou Yi Eight Trigrams is a role similar to a kind of dialect encryption, which can also improve the confidentiality of the algorithm [19].

There are many changing laws in the Zhou Yi Eight Trigrams. Combining these changing laws with chaotic system to encrypt images can greatly improve the security of the algorithm. Through the above description, combined with the background of offshore wind power, this paper proposes a new image encryption algorithm for the protection of offshore wind power image. The main innovation points of this paper are as follows.

(1) A new encryption rule is proposed based on the Zhou Yi Eight Trigrams, and the encryption and decryption algorithms are constructed by combining 2D-LCLM.

(2) Encrypt offshore wind power images and standard images. The experimental results show that the algorithm is effective in practical applications.

(3) Design comparative experiments (removing the Zhou Yi Eight). It is found that the Zhou Yi Eight Trigrams encryption rules can well improve the security performance of the algorithm.

The remainder of this paper is organized as follows. In Section 2, we give a brief introduction to the Zhou Yi and propose the Zhou Yi Eight Trigrams encryption rule. In Section 3, we propose a new encryption algorithm using 2D-LCLM, Sha-256, Zhou Yi Eight Trigrams and Zigzag. In Section 4, we adopt standard image and offshore wind images for experimental simulation, which prove that the algorithm has prefect encryption effect. We also design comparative experiments to verify the effect of Zhou Yi Eight Trigrams on the performances of the encryption algorithm. We provide a summary in Section 5.

**Tab.1** Obfuscation rule

| XOR result | 0 | 1 | 2 | 3 | 4 | 5 | 6 | 7 |
|---|---|---|---|---|---|---|---|---|
| Obfuscation result | Qian | Xun | Li | Gen | Dui | Kan | Zhen | Kun |

## 2 Zhou Yi Eight Trigrams
*2.1 Zhou Yi Eight Trigrams encryption rule*

The Zhou Yi Eight Trigrams is a basic philosophical concept widely used in ancient China and is composed of yin and yang lines. Arranging the yin and yang lines in any three directions from top to bottom, we can get the Zhou Yi Eight Trigrams (Qian☰, Kun☷, Zhen☳, Xun☴, Kan☵, Li☲, Gen ☶, Dui☱). Fig.1 shows the symbol of Zhou Yi Eight Trigrams. If the yang and yin lines are represented by 1 and 0, the binary representation of Zhou Yi Eight Trigrams are obtained (Qian 111, Kun 000, Zhen 001, Xun 110, Kan 010, Li 101, Zhen 100, and Dui 011).

**Fig.1** Zhou Yi Eight Trigrams symbol

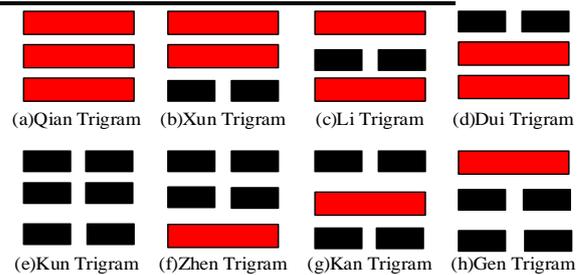

(a)Qian Trigram  (b)Xun Trigram  (c)Li Trigram  (d)Dui Trigram

(e)Kun Trigram  (f)Zhen Trigram  (g)Kan Trigram  (h)Gen Trigram

Combining the above Eight Trigrams, we can generate Ben Trigram, Zong Trigram, Cuo Trigram, Hu Trigram and other trigrams. They can be converted to each other in a certain way.

Hu Trigram is a new trigram composed of the fifth, fourth, third lines of the Ben Trigram as the upper trigram and the fourth, third, second lines as the lower trigram. Ben Trigram is turned over, then it becomes Zong Trigram. Cuo Trigram is a new trigram where the yin lines of the Ben Trigram are changed into yang lines, and yang lines into yin lines. For example, if one selects 101100 as Ben Trigram, then he can process 101100 to get the Hu Trigram 011110, the Zong Trigram 001101 and the Cuo Trigram 010011.

The Zhou Yi Eight Trigrams encryption rule employs the conversion method between Ben Trigram, Zong Trigram, Hu Trigram and Cuo Trigram to get another representation of the ciphertext, and then perform the XOR to get the result of Zhou Yi Eight Trigrams encryption rule. For instance, 101101 is encrypted by the Zhou Yi Eight Trigrams encryption rule, and we can get the encryption result 001100. The encryption process is shown in the Fig.2.

**Fig.2** Encryption process of the Zhou Yi Eight Trigrams encryption rule

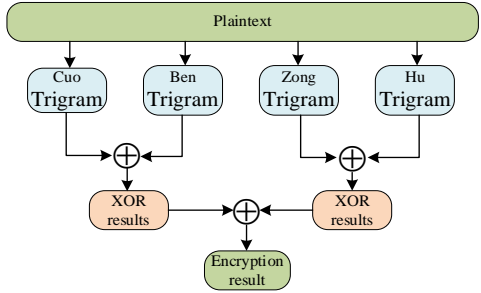

### 2.2 Confusion using the Zhou Yi Eight Trigrams

Before performing Feistel-like encryption, we first perform an obfuscation work on the plaintext data using gossip and secret key, and the specific process is as follows.

(1) We first give a random string of secret keys between 0 and 8, and convert each bit into three binary digits, noted as key3(i).

(2) Convert the decimal plaintext data into binary data, and then convert each bit into a group of three, noted as data3(i).

(3) Perform a XOR between key3 and data3. According to Tab.1, convert the XOR result into the corresponding Zhou Yi Eight Trigrams.

## 3. Encryption algorithms based on 2D-LCLM and Zhou Yi Eight Trigrams

### 3.1 Two-Dimensional (2D) Lag-Complex Logistic Map (LCLM)

The two-dimensional lagging complex logistic mapping (2D-LCLM) can be obtained by extending the two-dimensional logistic mapping from the real domain to the complex domain [11]. Compared with the two-dimensional (logistic) mapping, the chaotic characteristics of the 2D-LCLM is more complicated. Its definition is as follows.

$$\begin{cases} x_{n+1} = bx_n(1-z_n) \\ y_{n+1} = by_n(1-z_n) \\ z_{n+1} = ax_n^2 + y_n^2 \end{cases} \quad (1)$$

When a=1, b∈ [1.69,2), the mixed system is chaotic. Set a=1, b=1.99, the phase diagram of 2D-LCLM is as Fig.3.

**Fig.3** Phase diagram of 2D-LCLM

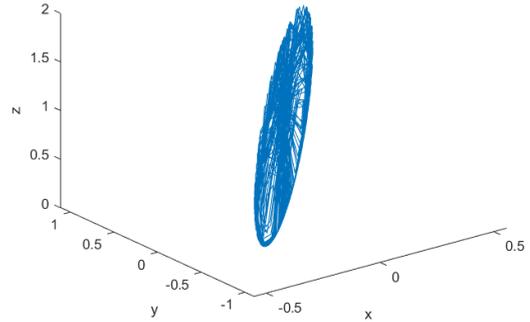

### 3.2 Feistel-like cipher structure

This article uses the Feistel-like cipher structure as the overall framework of the encryption algorithm. The round function $F$ is repeatedly executed in the Feistel-like cipher structure. The data is processed into a group of 48 bits, and each group is divided into 4 blocks $L_{0,1}$, $L_{0,2}$, $R_{0,1}$, $R_{0,2}$ equally. Perform $n$ rounds operations of Feistel-like cipher structure (Use the F function twice per round) and output $L_{r+1,1}$, $L_{r+1,2}$, $R_{r+1,1}$, $R_{r+1,2}$. The flow chart of the $F$ function is shown in Fig.4.

**Fig.4** Feistel-like cipher structure

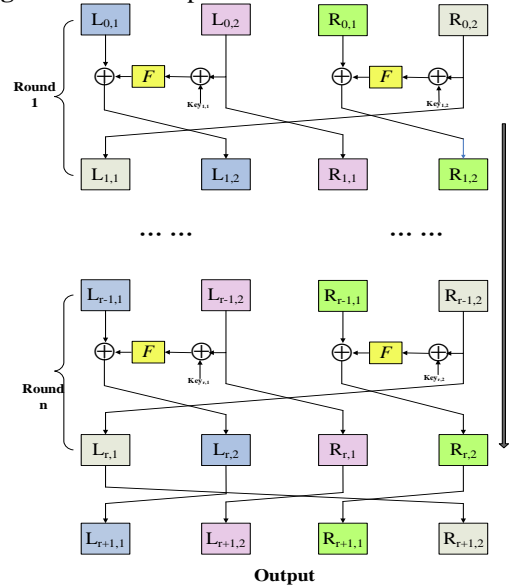

### 3.2.1 Construction of the F function

The task of obfuscation and diffusion is mainly undertaken by the $F$ function, the *Permutation*, and each round of position swapping in the Feistel-like cipher structure. Among them, the $F$ function is mainly composed of the S-box as well as the Zhou Yi Eight Trigrams encryption algorithm. The structure is shown in Fig.5.

The detailed process is as follows.

(1) First perform XOR operation on each input binary number (12bit) and round key.

(2) The XOR result is divided into two parts to be scrambled separately using S-box.

(3) The two parts of the output results are combined, and the position is scrambled using the QF function (QF is a nonlinear equation [20]. The transformation of the QF equation is shown in Tab.2).

242

(4) Finally, the 12-bit data is encrypted using the Zhou Yi Eight Trigrams encryption rule.

**Tab.2** QF function

| Kn, i∈1,2,3 | 0 | 1 | 2 | 3 | 4 | 5 | 6 | 7 | 8 | 9 | 10 | 11 |
|---|---|---|---|---|---|---|---|---|---|---|---|---|
| P (Kn, i∈1,2,3) | 5 | 2 | 10 | 6 | 7 | 4 | 0 | 1 | 3 | 8 | 9 | 11 |

*3.2.2 Composition of initial values of chaotic sequences*

(1) The data of the compressed image is noted as *plain*. Then calculate the decimal sum of *plain*, which is expressed by "*Sum_Pt*".

(2) The 256-bit hash value is obtained by the Sha-256 function [21] on "*Sum_Pt*", and the 256-bit hash value is set every 8 bits, $m_1, m_2, m_3, ..., m_{32}$. The initial values $x_1'(1)$, $x_2'(1)$, $x_3'(1)$ of the 2D-LCLM are calculated as follows,

$$\begin{cases} x_1'(1) = x_1(1) + \mod(\frac{m_1 \oplus ... \oplus m_8 + mean}{256}, 1) \\ x_2'(1) = x_2(1) + \mod(\frac{m_9 \oplus ... \oplus m_{16} + mean}{256}, 1) \\ x_3'(1) = x_3(1) + \mod(\frac{m_{17} \oplus ... \oplus m_{24} + mean}{256}, 1) \\ mean = \sum_{i=1}^{32} \frac{m_i}{32} \end{cases} \quad (2)$$

where $x_1(1)$, $x_2(1)$, $x_3(1)$ are used as the original initial values of the 2D-LCLM. Therefore, the chaotic iteration and the plaintext are linked.

**Fig.5** Structure of the round function *F*

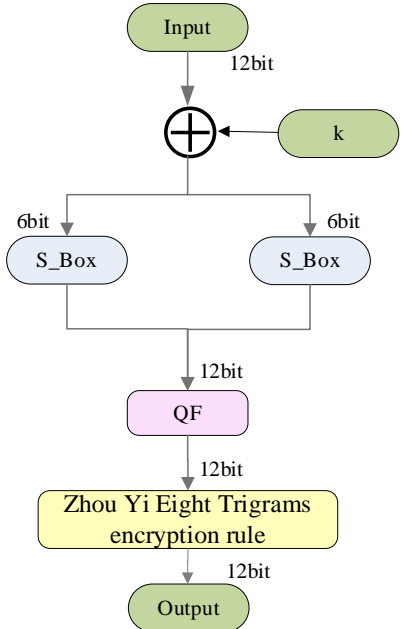

*3.2.3 Construction of position sequences and integer sequences*

According to the previous steps, the initial values of 2D-LCLM are obtained, and the 2D-LCLM chaotic system iterates to obtain the pseudo-random sequences $X_1$, $X_2$ and $X_3$.

(1) Construction of the position sequence

A pseudo-random sequence is selected and the numbers in the sequence are arranged from the smallest to the largest. By recording the position of each number in the original random sequence, a position sequence *T* is obtained.

(2) Construction of integer sequence

Each element of the random sequence $X_i$ ($i$=1,2,3) is transformed into an integer between (0,255) or (0,63) to obtain the integer sequence $Z_1$ or $Z_2$.

$$Z_1 = floor(((X_i \times 10^3 - floor(X_i \times 10^3)) \times 10^3)) \mod 256 \quad (3)$$

$$Z_2 = floor(((X_i \times 10^3 - floor(X_i \times 10^3)) \times 10^3)) \mod 64$$

*3.2.4 Generation of the S_box*

The S_box is generated by pseudo-random sequence and Zigzag permutation [22]. Fig.6 is its flowchart.

Firstly, obtain the initial values of 2D-LCLM by processing the plaintext data, iterate the 2D-LCLM and get the pseudo-random sequences $X_1$, $X_2$ and $X_3$.

Secondly, a string of consecutive real numbers is selected from the obtained pseudo-random sequence $X_2$ and obtain the non-repeating real numbers between (0,63) by modulo and rounding operations on these real numbers, then get an S1_box.

Finally, get an S2_box through scrambling according to the position sequence generated from $X_2$. Dislocate the S2_box through the Zigzag to generate the final S_box.

**Fig.6** Structure of the S-box

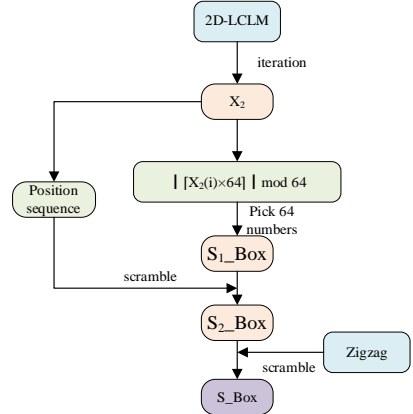

*3.2.5 Construction of the round key*

The round key is generated by using the Zhou Yi Eight Trigrams encryption rule, and it will change with the plaintext. The detailed process is as follows.

(1) Give a random string of hexadecimal initial secret key and convert each bit into binary.

(2) Scramble the binary sequence according to the position sequence generated from the pseudo-random sequence.

(3) Divide every six bits into a group in the shuffled binary sequences. And encrypt it by using the Zhou Yi Eight Trigrams encryption rule to generate the round key. The construction of the round key can be expressed by the following formula.

$$k = ZY\_rule(T(hex2bin(key, 4))) \quad (4)$$

where *ZY_rule* is Zhou Yi Eight Trigrams encryption rule, *T* indicates position scrambling, *hex2bin* represents the conversion of hexadecimal numbers to binary.

*3.2.6 Permutation*

To enhance the pseudo-randomness of binary data, in addition to the Feistel algorithm scrambling, *Permutation* is used before and after the Feistel algorithm. The structure of *Permutation* is shown in Fig.7.

**Fig.7** Structure of Permutation

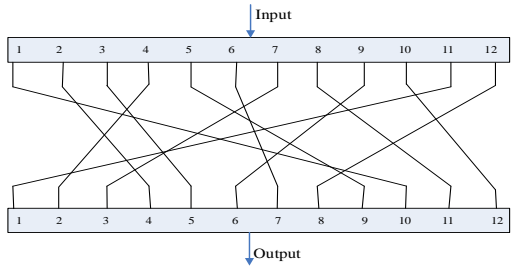

*3.3 Encryption algorithm*

We divide the plaintext image into R, G, B three channels and perform the same operation. Fig.8 shows the encryption flow chart of R channel. The whole encryption process is described in detail.

Step 1: Read the image to be encrypted and compress it to the specified size to get the compressed image "img".

Step 2: The R, G, and B channels in "img" are separated. Read the data of R channel and write it down as "Pt". An obfuscation work is performed on "Pt" to obtain the image "Pt1" (As shown in subsection 3.2.3).

Step 3: Perform a decimal summation operation on all "Pt1", the result is recorded as "Sum_Pt1".

Step 4: Use the Sha-256 to calculate the 256-bit hash value of "Sum_Pt1" and obtain the initial values $x_1'(1)$, $x_2'(1)$, $x_3'(1)$ of 2D-LCLM by subsection 3.2.2.

Step 5: The initial values obtained in the previous steps are brought into the 2D-LCLM complex chaotic system to obtain three pseudo-random sequences $X_1$, $X_2$ and $X_3$, which are used to construct a sequence of integers and a sequence of positions (As shown in subsection 3.2.3).

Step 6: Use the random sequence $X_2$ to construct the position sequence $T_1$. Get the 64 non-repeating numbers between (0,63) from $X_2$, and obtain the initial S-box by modulo integer operation. Through Zigzag permutation and position sequence $T_1$, a dynamic S-box is obtained. (As shown in subsection 3.2.4)

Step 7: The position sequence $T_2$ is constructed from the random sequence $X_2$. A string of initial hexadecimal secret keys is given in advance and converted into binary numbers. The binary round key $K$ is obtained by using Zhou Yi Eight Trigrams encryption rule (As shown in section 2) and position sequence $T_2$ (As shown in subsection 3.2.5).

Step 8: Firstly, read the data of the compressed image "img" and obtain P1 by limiting all its values to (0,255). Secondly, use the random sequence $X_2$ to construct the position sequence $T_3$, and employ $T_3$ disrupt P1 to obtain P2, P2 = $T_3$ (P1); Then, adopt the random sequence $X_1$ to construct the position sequence $T_4$, and use $T_4$ disrupt P2 to obtain P3, P3 = $T_4(P2)$; Finally, each bit in P3 is changed into binary to get P4. P4 is divided into a group of 48bit recorded as "plain_block(i)".

Step 9: Encrypt each group of data by Feistel-like cipher structure. The output permutation of the last round is: $L_{j+1,1}=L_{j,2}$, $L_{j+1,2}=R_{j,1}$, $R_{j+1,1}=R_{j,2}$, $R_{j+1,2}=L_{j,1}$. Then link the output ciphertext "Ciphered_block(i)" through Cipher-block chaining (CBC) to get the final binary cipher "Ciphered_data".

Step 10: Convert each octet of binary into an integer, and finally translate it into the final ciphertext by ASCII.

Step 11: The same operation is performed on channels G and B to obtain their ciphertexts. Finally, the three channels are combined to get the cipher-image.

**Fig.8** Encryption flow chart of R-channel

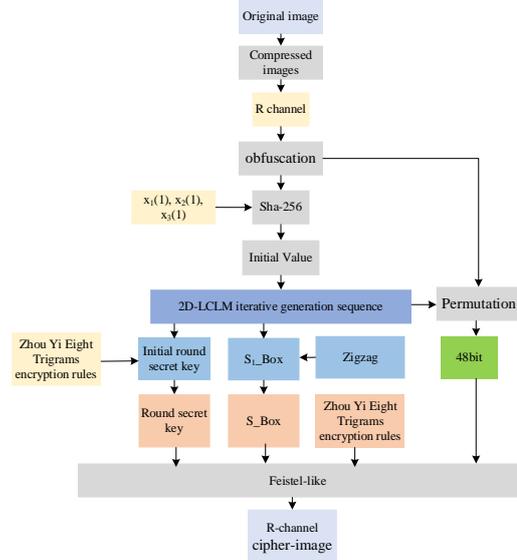

*3.4 Decryption algorithm*

The decryption process is the inverse of the encryption process.

(1) Read the cipher-image and convert it to binary sequence.

(2) The binary sequence is divided into several groups of 48 bits, which are encrypted using a Feistel-like cipher structure. The data permutation in the last round of output is different from encryption and becomes $L_{j+1,1}=R_{j,2}$, $L_{j+1,2}=L_{j,1}$, $R_{j+1,1}=R_{j,2}$, $R_{j+1,2}=R_{j,1}$.

(3) Link the output and get binary data of the compressed image by CBC.

(4) Convert the binary data into integers in groups of eight bits and perform an inverse process of obfuscation to obtain the compressed image.

**4 Experimental results and analysis**

*4.1 Simulation of standard image and offshore wind power images*

We set $x_1(1) = 0.2$, $x_2(1) = 0.4$, $x_3(1) = 0.1$ and select the standardized test image, the wind turbine image and the substation image (which will be shown in Case1, Case2 and Case3). First the pixel of these images is compressed to 240×240. Then we use the encryption algorithms to get the encrypted images that cannot be recognized by the naked eye (As shown in Fig.9). In the following, we test the algorithm in detail by histogram, information entropy, pixel correlation, etc.

**Fig 9.** Encryption and decryption image. (a), (d) and (g) are compressed images of Case1, Case2, and Case3, respectively. (b), (e), (h) are cipher-images of (a), (d) and (g), respectively. (c), (f), (i) are decrypted images of (b), (e), (h), respectively.

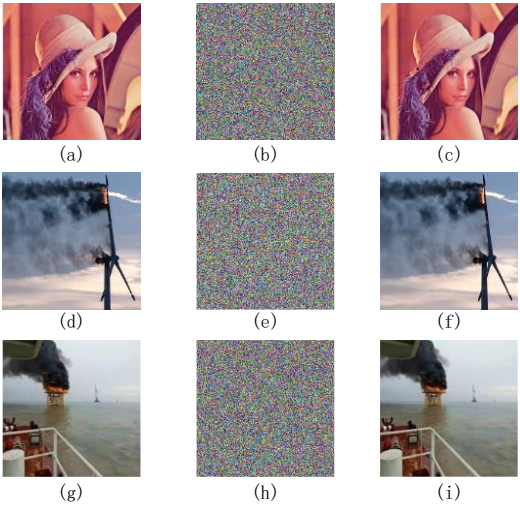

### 4.1.1 Histogram analysis

The histogram allows us to determine whether the pixel values of an image are evenly distributed. Fig.10 is the histogram before and after encryption in R, G, and B channels of "Case1". We can see that the pixels are evenly distributed in each gray level after encryption, which can hide the information of the image well and resist statistical attacks.

**Fig.10** Histogram of the Case1. (a) R channel in plain image; (b) G channel in plain image; (c) B channel in plain image; (d) R channel in cipher image; (e) G channel in cipher image; (f) B channel in cipher image.

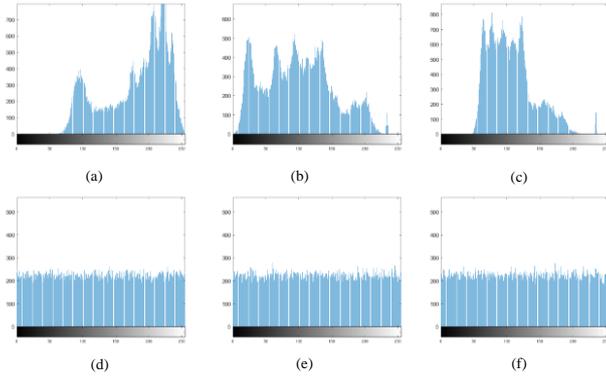

### 4.1.2 Information entropy

Information entropy can be used to measure the degree of data orderliness in the information source. The closer the information entropy is to 8, the more disorderly the data in the information source is, and the better the corresponding image encryption effect is. The formula for the information entropy H is as follows,

$$H = -\sum_{i=1}^{L} p(x_i) \log_2 p(x_i) \quad (5)$$

where $L=256$, $p(x_i)$ is the probability of occurrence of $x_i$ in the ciphertext.

From the results in Tab.3, it can be seen that the information entropy of the image has been greatly improved after encryption. Compare the information entropy of this paper and the literature [23]. It is found that the information entropy of this paper is slightly better than [23], which means that our algorithm is more effective.

### 4.1.3 Correlation test

As there is a strong correlation between adjacent pixels, we can see the information in an image. Encrypting an image is to disrupt this correlation. The experiments are randomly selected $N=8000$ pairs of adjacent points from horizontal, vertical and diagonal directions for Case1, Case2 and Case3. Each pair of pixel points is denoted by $(u_i,v_i)$, $i \in [1,N]$. The formula of the correlation $r_{u,v}$ between $u_i$ and $v_i$ as follows.

$$\begin{cases} E(u) = \dfrac{1}{N}\sum_{i=1}^{N} u_i \\ D(u) = \dfrac{1}{N}\sum_{i=1}^{N}(u_i - E(u))^2 \\ Cov(u,v) = \dfrac{1}{N}\sum_{i=1}^{N}(u_i - E(u))(v_i - E(u)) \\ r_{u,v} = \dfrac{Cov(u,v)}{\sqrt{D(u)} \times \sqrt{D(v)}} \end{cases} \quad (6)$$

Fig.11 shows the result for the R channel of the wind turbine image. Fig.11 (a)-(c) is the correlation of the original image in the vertical, horizontal and diagonal directions, where the pixel points of the plaintext image are clearly correlated in all three directions and the correlation tends to 1. Fig.11(d)-(f) is the correlation of the cipher image in all three directions, where the pixel points are almost equally distributed, and the correlation tends to 0. Tab.4 records the correlation data of Case1, Case2 and Case3. Compare them with the correlation data of the references [23-25]. It is found that the correlation data obtained after encryption are mostly smaller than those of the literature. Therefore, the pixels are sufficiently disrupted after the encryption.

**Tab.3** Information entropy of images

| Image | Plain image | | | Cipher image | | |
|---|---|---|---|---|---|---|
| | R | G | B | R | G | B |
| Case1 | 7.4003 | 7.3885 | 7.3955 | 7.9968 | 7.9969 | 7.9968 |
| Case2 | 7.6346 | 7.6084 | 7.3207 | 7.9971 | 7.9969 | 7.9963 |
| Case3 | 7.4726 | 7.6503 | 7.7509 | 7.9967 | 7.9968 | 7.9966 |
| Ref. [23] | 7.2796 | 7.6321 | 6.9892 | 7.9973 | 7.9969 | 7.9971 |

**Tab.4** Correlation coefficients in adjacent pixels before and after

| Image | Plain image | | | Cipher image | | |
|---|---|---|---|---|---|---|
| | Horizontal | Vertica | Diagonal | Horizontal | Vertica | Diagonal |
| Case1_R | 0.9449 | 0.9779 | 0.9224 | -0.0054 | 0.0216 | 0.0014 |
| Case1_G | 0.9143 | 0.9724 | 0.8868 | 0.0029 | 0.0122 | -0.00079 |
| Case1B | 0.8809 | 0.9568 | 0.8584 | 0.0059 | 0.0065 | -0.0175 |
| Case2_R | 0.9655 | 0.9895 | 0.9634 | 0.0100 | -0.0206 | -0.0047 |
| Case2_G | 0.9512 | 0.9843 | 0.9538 | 0.0219 | -0.0096 | -0.0227 |
| Case2_B | 0.9359 | 0.9840 | 0.9400 | -0.0082 | 0.0112 | -0.0010 |
| Case3_R | 0.9829 | 0.9768 | 0.9752 | -0.0059 | 0.0036 | 0.0192 |
| Cas 3_G | 0.9878 | 0.9792 | 0.9794 | 0.0058 | -0.0092 | 0.0173 |
| Case3_B | 0.9889 | 0.9804 | 0.9789 | -0.0155 | 0.0216 | -0.0079 |
| Ref. [23] | 0.9618 | 0.9359 | 0.9001 | -0.0029 | 0.0013 | -0.0026 |
| Ref. [24] | 0.9668 | 0.9416 | 0.8973 | -0.0223 | -0.0084 | -0.0086 |
| Ref. [25] | 0.9964 | 0.9988 | 0.9950 | 0.0693 | 0.0610 | -0.0242 |

**Tab.5** NPCR and UACI values of ciphered images

| Image | NPCR(%) | | | UACI(%) | | |
|---|---|---|---|---|---|---|
| | R | G | B | R | G | B |
| Lena | 99.8507 | 99.4531 | 99.8073 | 33.6227 | 33.4879 | 33.4959 |
| Case2 | 99.4688 | 99.3715 | 98.9913 | 33.4943 | 33.3328 | 33.5626 |
| Case3 | 99.4566 | 99.9080 | 99.4583 | 33.3616 | 33.4649 | 33.4146 |
| Ref. [23] | 99.6000 | 99.6100 | 99.6100 | 33.5600 | 33.4500 | 33.4900 |
| Ref. [24] | 99.5697 | 99.6094 | 99.5789 | 33.4100 | 33.4635 | 33.4409 |
| Ref. [25] | | 99.5544 | | | 33.4549 | |

**Fig.11** Distribution map of the adjacent pixels in wind turbine. (a)Plain image Vertical pixel distribution; (b)Plain image Horizontal pixel distribution; (c)Plain image Diagonal pixel distribution; (d)Cipher image Vertical pixel distribution; (e)Cipher image Horizontal pixel distribution; (f)Cipher image Diagonal pixel distribution.

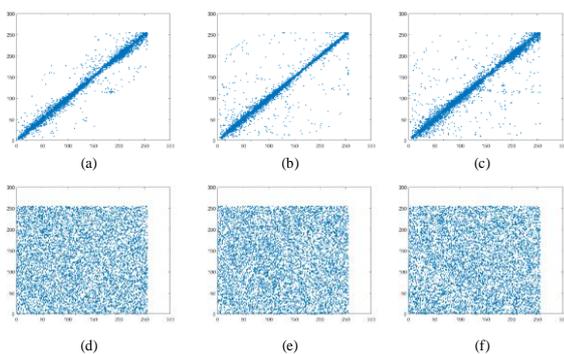

### 4.1.4 Plaintext sensitivity test

The plaintext sensitivity test can clearly show the difference between the output ciphertext and the original ciphertext when the pixel value in the plaintext is changed. The UACI and NPCR are important indicators to measure the sensitivity of plaintext. The mathematical expressions of NPCR and UCAI are as follows,

$$\begin{cases} NPCR = \dfrac{\sum_{i,j}^{M,N} D(i,j)}{M \times N} \times 100\% \\ UACI = \dfrac{\sum_{i,j}^{M,N} |C_1(i,j) - C_2(i,j)|}{M \times N \times 255} \times 100\% \\ D(i,j) = \begin{cases} 1, (C_1(i,j) \neq C_2(i,j)) \\ 0, (C_1(i,j) = C_2(i,j)) \end{cases} \end{cases} \quad (7)$$

where $M$ and $N$ denote the length and width of the image, $C_1$ and $C_2$ is the pixel values of the cipher image before and after the change of a single pixel point, and $(i,j)$ denotes the position of the pixel point in the image.

From the results in Tab.5, both NPCR and UACI are close to the ideal values and mostly better than the references [23-25]. Therefore, our algorithm has a better ability to resist differential attacks and has good robustness.

### 4.1.5 Key space analysis

The secret key space is the range of the secret key size, which should be generally larger than $2^{100}$. In our paper,

**Tab.6** PSNR under noise attack

| Image | SPN 0.001 | SPN 0.005 | SN 0.000002 | SN 0.000005 | GN 0.000001 | GN 0.000003 |
|---|---|---|---|---|---|---|
| Case3_R | 22.2294 | 16.3571 | 30.4598 | 19.7496 | 21.7997 | 15.7303 |
| Case3_G | 22.5750 | 15.3029 | 28.7853 | 18.3672 | 21.0330 | 15.1524 |
| Case3_B | 21.8270 | 15.7069 | 25.5203 | 17.4226 | 19.3891 | 14.1383 |

the secret key consists of three parts: (1) the 256-bit hash value generated by Sha-256; (2) the parameters *a*, *b* and the initial value $x_1(1)$, $x_2(1)$, $x_3(1)$ of 2D-LCLM; (3) a string of hexadecimal round key and a string of decimal secret key. Assume accuracy of the computer memory is $10^{15}$, the secret key space in the 2D-LCLM chaotic system is $10^{15 \times 5}$ much larger than $2^{100}$. Together with the 256-bit hash and the secret key, it is enough to prove that the secret key space of the algorithm can resist against all kinds of brute force attacks.

*4.1.6 Cropping attacks and Noise attack analysis*

Images are subject to clipping attacks and noise attacks when transmitted over a common channel, and a good algorithm should be robust to such attacks. This section will analyze this algorithm in these two aspects.

The Peak Signal-to-Noise Ratio (PSNR) is used to quantitatively evaluate the quality of the decrypted image. The mathematical expression of PSNR is as follows,

$$\begin{cases} PSNR = 10\log_{10}(\frac{MAX^2}{MSE}) \\ MSE = \frac{1}{M \times N} \sum_{i=1}^{N}\sum_{j=1}^{M} \|K(i,j) - I(i,j)\|^2 \end{cases} \quad (8)$$

where $M \times N$ denotes the size of the image, $K(i,j)$ and $I(i,j)$ is the pixel values of the image before and after the attack, *MAX*=255, and *MSE* is the mean square error between *K* and *I*. When the calculated *PSNR* is smaller, the more serious the distortion of the image is.

(1) Noise attack analysis

In order to verify whether the algorithm can resist the noise attack, this paper adds Gaussian noise (GN), Salt & Pepper noise (SPN) and Speckle noise (SN) with different noise densities to the cipher image. Fig.12 shows the decryption image with different noise, and Tab.6 is the *PSNR* of the decrypted graph under different noise attacks. From the experimental results, we can see that after adding noise to the cipher image, the recovered image can still show the basic information of the original image, which proves that the algorithm has certain anti-noise capability in transmitting image data of offshore wind farm.

**Fig.12** Decrypted graph after noise attack

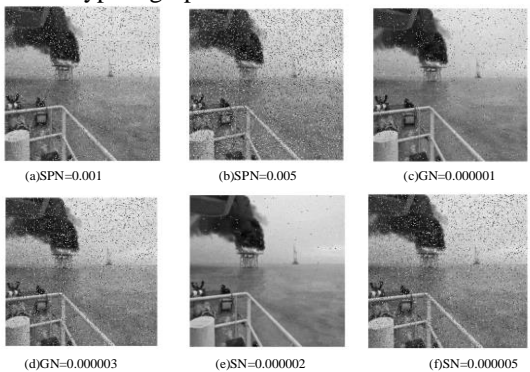

(a)SPN=0.001  (b)SPN=0.005  (c)GN=0.000001
(d)GN=0.000003  (e)SN=0.000002  (f)SN=0.000005

(2) Cropping attacks analysis

In this part, the cipher images are cropped 48*48 pixels, 60*60 pixels, and 96*96 pixels. Fig.13 shows the cipher images subjected to cropping and their decrypted images. Tab.7 shows the *PSNR* of the decrypted image after the cropping attack. As the cropped pixels become larger, the *PSNR* grows smaller and the decrypted image becomes more and more blurred, but the cipher image can still recover the basic information. It proves that the algorithm can resist the cropping attack.

**Fig.13** Cipher image under clipping attack and its decrypted image

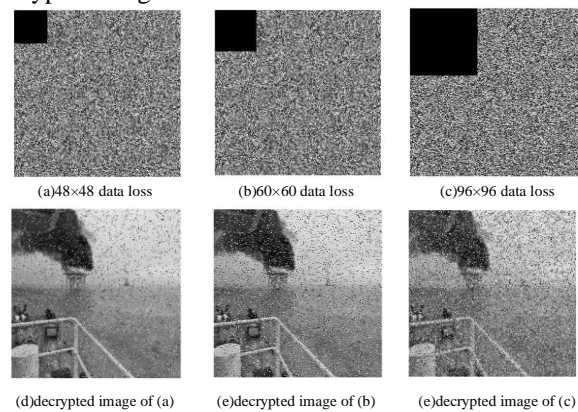

(a)48×48 data loss  (b)60×60 data loss  (c)96×96 data loss

(d)decrypted image of (a)  (e)decrypted image of (b)  (e)decrypted image of (c)

**Tab.7** PSNR under Cropping attacks

| image | 48×48 | 60×60 | 96×96 |
|---|---|---|---|
| Case3_R | 18.3832 | 16.8876 | 13.8868 |
| Case3_R | 17.9693 | 16.4566 | 13.5199 |
| Case3_R | 17.6489 | 16.2419 | 13.2334 |

*4.2 Analysis of the Zhou Yi Eight Trigrams encryption rule*

To verify the effectiveness of the Zhou Yi Eight Trigrams encryption rule, we remove the Zhou Yi Eight Trigrams encryption rule from our algorithm (call it N-ZY) and compare its performance with the full algorithm (call it ZY).

Tab.8 shows the performances comparison of the two algorithms on information entropy, NPCR and UACI. Compared with N-ZY, ZY algorithm has some improvements on the above performances, which indicates that incorporating Zhou Yi Eight Trigrams into the encryption algorithm can improve the security performances of the algorithm.

**Tab.8** Comparison of performances

| Methods | NPCR | UACI | Entropy |
|---|---|---|---|
| ZY | 99.7256 | 33.5355 | 7.9969 |
| N-ZY | 99.4774 | 33.4144 | 7.9967 |

**5.Conclusions**

In this paper, a new encryption algorithm is proposed based on Zhou Yi Eight Trigrams encryption rule and the 2D-LCLM. In the algorithm, the processed plaintext is the input of Sha-256, and its output is used to generate the initial values of the 2D-LCLM chaotic system. The round key is constructed by the Zhou Yi Eight Trigrams encryption rule; the S-box is composed of the sequence of integers generated by the 2D-LCLM chaotic system and Zigzag. We adopt Case1, Case2 and Case3 to encrypt and make experiments. The results show that the algorithm is better than N-ZY in entropy, NPCR and UACI, and it also can resist noise attacks, clipping attacks, etc.